\documentclass[twocolumn,showpacs,preprintnumbers,amsmath,amssymb,twoside]{revtex4}
\usepackage[latin1]{inputenc}
\usepackage{bbm,hyperref}

\newcommand{\cc}{\text{\sc c}}
\newcommand{\gut}{\text{\sc gut}}
\newcommand{\hbcc}{\overline{H}_{\cc}}
\newcommand{\hbf}{\overline{H}_{\!f}}
\newcommand{\sm}{\text{\sc sm}}

\DeclareMathOperator{\diag}{diag}

\newcommand{\bt}[1]{b^{\dagger}_{#1}}
\newcommand{\ket}[1]{\left| #1\right\rangle}              % | >
\newcommand{\braket}[3]{\left\langle #1\!\left| #2\right|\!
    #3\right\rangle} % < | | > 
\newcommand{\VEV}[1]{\left\langle #1\right\rangle}        % < >

\begin{document}

\preprint{hep-ph/0501223}

\preprint{ILL-(TH)-05-01}

\title{Operator Analysis for Proton Decay in SUSY {\sffamily SO(10)}
  GUT Models}

\author{S\"oren Wiesenfeldt}
%\email{soeren@uiuc.edu}
\affiliation{Department of Physics, University of Illinois at
  Urbana-Champaign, 1110 West Green Street, Urbana, IL 61801,
  USA}

\date{January 24,2005}

\begin{abstract}
  \noindent
  Non-renormalizable operators both account for the failure of down
  quark and charged lepton Yukawa couplings to unify and reduce the
  proton decay rate via dimension-five operators in minimal SUSY {\sf
    SU(5)} GUT.  We extend the analysis to SUSY {\sf SO(10)} GUT
  models.
\end{abstract}

\pacs{12.10.-g; 12.60.Jv}

\maketitle

%%%%%%%%%%%%%%%%%%%%%%%%%%%%%%%%%%%%%%%%%%%%%%%%%%%%%%%%%%%%%%%%%%%%%%%%

\noindent
The standard model (SM) is a very successful but only effective
theory, which has to be extended at higher energies.  Grand Unified
Theories provide a beautiful framework for a more fundamental theory
because the additional symmetries of the underlying group
$\mathsf{G_\gut}$ restrict some of the arbitrary features, in
particular via relations between Yukawa couplings.  Unfortunately,
these relations are troublesome in simple GUT models like minimal {\sf
  SU(5)} \cite{georgi74}: The equality $m_d=m_e$ is correct for the
third but fails for the first and second generation.  It was, however,
noticed early that since the GUT scale $M_\gut$ is close to the Planck
scale $M_\text{P}$, it is natural to expect corrections to the fermion
mass matrices, which do not respect this equality \cite{ellis79b}.
Moreover, these corrections are important for supersymmetric GUT
models because they can significantly reduce the proton decay rate via
dimension-five operators.  For minimal supersymmetric {\sf SU(5)}
\cite{susy-su5}, they are sufficient to make it consistent with the
present experimental bound \cite{cw03,bajc}.

Since these additional contributions are important, it is reasonable
to study them in more detail. In particular, we would like to know,
whether they are totally arbitrary from the GUT model's point of view,
or whether there is any mechanism which would naturally lead to the
required relations among Yukawa couplings.  We can think of two
possibilities, the first of which is to start with some ad-hoc
textures as a result of an unknown additional symmetry
\cite{textures}; the second is to extend the analysis to another
group, in order to obtain additional symmetry restrictions.

For the latter approach, {\sf SO(10)} \cite{so10} is a natural choice
since it unifies the matter fields and and involves massive neutrinos.
It is broken to the SM either via the Pati-Salam group \cite{pati74}
or $\mathsf{SU(5)\times U(1)}$.  Higher-dimensional operators have
been studied in the former case before (see e.\,g.
Refs.~\cite{babu98,aulakh}); the purpose of this letter is to study
the non-renormalizable operators for the second case.  Here, these
operators have been studied in parts only, see e.\,g.
Ref.~\cite{chang04,nath-syed}.

\smallskip

%====================================================================

At the beginning, we review the higher-dimensional operators in
supersymmetric {\sf SU(5)}.  The minimal model contains three
generations of chiral matter multiplets, \mbox{$\mathsf{10} =
  (Q,u^\cc,e^\cc)$}, \mbox{$\mathsf{5}^* = (d^\cc,L)$}, and as Higgs
fields, an adjoint multiplet $\Sigma(\mathsf{24})$ and a pair of
quintets, $H(\mathsf{5})$ and $\overline{H}(\mathsf{5}^*)$.  $\Sigma$
acquires the vacuum expectation value (vev) \mbox{$\VEV{\Sigma}=\sigma
  \diag \left(2,2,2,-3,-3\right)$}, where $\sigma\simeq M_\gut$, so
that {\sf SU(5)} is broken to ${\sf G_\sm}$.  The pair of quintets
then breaks ${\sf G_\sm}$ to ${\sf SU(3)\times U(1)_\text{em}}$ at
$M_\text{ew}$; it contains the SM Higgs doublets, $H_f$ and $\hbf$,
which break $\mathsf{G_\sm}$, and color triplets, $H_\cc$ and $\hbcc$,
respectively.
Including possible terms up to order $1/M_\text{P}$, the
Yukawa couplings read
\begin{align} \label{eq:consistent-w}
  W & = \frac{1}{4}\,\epsilon_{abcde} \left( Y_1^{ij}
    \mathsf{10}_i^{ab} \mathsf{10}_j^{cd} H^e \right.
  \nonumber \\
  & \qquad + \left. f_1^{ij} \mathsf{10}_i^{ab} \mathsf{10}_j^{cd}
    \frac{\Sigma^e_f}{M_\text{P}} H^f + f_2^{ij} \mathsf{10}_i^{ab}
    \mathsf{10}_j^{cf}\, H^d \frac{\Sigma^e_f}{M_\text{P}} \right)
  \nonumber \\
  & \quad + \sqrt{2} \left( Y_2^{ij} \overline{H}_{\!a} \mathsf{10}_i^{ab}
    \mathsf{5}^*_{jb} \right.
  \nonumber \\
  & \qquad + \left. h_1^{ij} \overline{H}_{\!a}
    \frac{\Sigma^a_b}{M_\text{P}} \mathsf{10}_i^{bc} \mathsf{5}^*_{jc}
    + h_2^{ij} \overline{H}_{\!a} \mathsf{10}_i^{ab}
    \frac{\Sigma_b^c}{M_\text{P}} \mathsf{5}^*_{jc} \right) ,
\end{align}
thus
\begin{align}
  \label{eq:sm-planck} 
  \begin{split} 
    Y_u & = Y_1 + 3\frac{\sigma}{M_\text{P}}f_1^S +
    \frac{1}{4}\frac{\sigma}{M_\text{P}}\left(3f_2^S+5f_2^A\right)
    \ , \\[2pt]
    Y_d & = Y_2 - 3\frac{\sigma}{M_\text{P}}h_1 +
    2\frac{\sigma}{M_\text{P}}h_2
    \ , \\[2pt]
    Y_e & = Y_2 -3\frac{\sigma}{M_\text{P}}h_1 -
    3\frac{\sigma}{M_\text{P}}h_2 \ .
  \end{split}
\end{align} 
Here $\sigma/M_\text{P}\sim{\cal O}\left(10^{-2}\right)$, and $S$ and
$A$ denote the symmetric and antisymmetric parts of the matrices.
From Eqs.~(\ref{eq:sm-planck}) one reads off,
\begin{align}
  \label{eq:c-relation}
  Y_d - Y_e = 5\frac{\sigma}{M_\text{P}}h_2 \ ,
\end{align}
hence the failure of Yukawa unification is naturally accounted for by
the presence of $h_2$.  

The dimension-five operators that lead to proton decay arise from the
couplings of quarks and leptons to $H_\cc$ and $\hbcc$, which acquire
masses ${\cal O}\left(M_\gut\right)$. If we integrate them out, two
operators remain \cite{dim5op},
\begin{align} \label{eq:operator}
  W_5 = \frac{1}{M_{H_\cc}} \left[ \tfrac{1}{2} Y_{qq} Y_{ql}\, QQQL +
    Y_{ue} Y_{ud}\, u^\cc e^\cc u^\cc d^\cc \right] ,
\end{align}
where 
\begin{align}
  \begin{split}
    Y_{qq} = Y_{qq}^S = Y_{ue}^S &= Y_u^S -
    5\frac{\sigma}{M_\text{P}}\left( f_1^S+\frac{1}{4} f_2^S\right) ,
    \\[2pt]
    Y_{ue}^A & = Y_u^A - \frac{5}{2}\,\frac{\sigma}{M_\text{P}}f_2^A
    \ , 
    \\[2pt]
    Y_{ql} & = Y_e + 5\frac{\sigma}{M_\text{P}}h_1 \ ,  
    \\[2pt]
    Y_{ud} & = Y_d + 5\frac{\sigma}{M_\text{P}}h_1 \ .
  \end{split}
\end{align}
The entries in $f_j$ and $h_j$ can lead to a simple pattern of these
Wilson coefficients with small entries only \cite{cw03}.

\medskip

%====================================================================

In {\sf SO(10)}, the analogous five-dimensional operator is ${\sf 16\ 
  16\ 10}_H\, {\sf 45}_H$.  
Here, we use the scenario, where {\sf SO(10)} is broken to {\sf SU(5)}
by a pair ${\sf 16}_H \oplus {\sf 16}^*_H$ (cf. the discussion at the
end of this letter). {\sf SU(5)} is broken to $\mathsf{G_\sm}$ by the
adjoint ${\sf 45}_H$ which includes the $\Sigma(\mathsf{24})$ of {\sf
  SU(5)}; finally the breaking of $\mathsf{G_\sm}$ is achieved by
${\sf 10}_H$ which includes both $H(\mathsf{5})$ and
$\overline{H}(\mathsf{5}^*)$ of {\sf SU(5)}.  This breaking pattern
guarantees that the {\sf SO(10)} gauge coupling constant remains
perturbative up to $M_\text{P}$ \cite{chang04}.

In order express the {\sf SO(10)} in {\sf SU(5)} fields, we consider a
set of operators $b_j\ (j=1,\dots, 5)$ plus their Hermitean
conjugates, $b_j^\dagger$, satisfying \cite{mohapatra80}
\begin{align}
  \left\{ b_i, b_j \vphantom{\bt{j}}\right\} & = \left\{ \bt{i},
    \bt{j}\right\} = 0 \ , & \left\{ b_i, \bt{j}\right\} & =
  \delta_{ij} \ .
\end{align}
With $\Gamma$ matrices, defined as 
\begin{align}
  \Gamma_{2j-1} & = -i \left(\, b_j - \bt{j}\, \right) , & 
  \Gamma_{2j} & = \left(\, b_j + \bt{j}\, \right) , 
\end{align}
we can construct the generators of {\sf SO(10)}, $\Sigma_{\mu\nu}$, as
\begin{align}
  \Sigma_{\mu\nu} = \frac{1}{2i} \left[ \Gamma_\mu, \Gamma_\nu \right]
  \ .
\end{align}
The spinor representation can be split into two 16-dimensional
representations $\Psi_\pm$ by chiral projection, $\tfrac{1}{2} \left(
  1 \pm \Gamma_0 \right)$, $\Gamma_0=i\prod_j\Gamma_j$.  We define an
{\sf SU(5)} invariant vacuum state $\ket{0}$ and expand the spinors in
terms of {\sf SU(5)} fields.  The SM fermions are assigned to {\sf
  16},
\begin{align} \label{eq:16-dec}
  {\sf 16} = \ket{\Psi_+} & = \ket{0}\,\psi_0 +
  \frac{1}{2!}\,\bt{i}\bt{j}\ket{0}\psi^{ij}
  \nonumber \\[2pt]
  & \qquad + \frac{1}{4!}\,\epsilon^{ijklm}\,
  \bt{j}\bt{k}\bt{l}\bt{m}\ket{0}\,\overline{\psi}_{i} \ ,
\end{align}
where we identify $\overline{\psi}_{i}$ and $\psi^{ij}$ with $5^*$ and $10$
of {\sf SU(5)}, respectively.  The singlet $\psi_0$ denotes the
left-handed anti-neutrino.

%====================================================================

The fundamental representation can be written in {\sf SU(5)} fields as
\begin{align} \label{eq:so10-10H}
  \phi_\mu & = \left\{
    \begin{array}{ll}
      \phi_{2j}\!\!\! & = \frac{1}{2}\, \left(
        \phi_{c_j} + \phi_{\bar c_j} \right) \\[1mm]
      \phi_{2j-1}\!\!\! & = \frac1{2i} \left(
        \phi_{c_j} - \phi_{\bar c_j} \right)
    \end{array}
  \right. \ ,
\end{align}
where $\phi_{c_j}$ and $\phi_{\bar c_j}$ transform like {\sf SU(5)}
representations.  Thus we are able to compute the {\sf SO(10)} in {\sf
  SU(5)} fields which then only have to be reduced to irreducible
representations.  We obtain
\begin{align} \label{eq:so10-gamma10}
  \Gamma_\mu\, \phi_\mu
  & = b_j\, \phi_{c_j} + \bt{j}\, \phi_{\bar c_j} \ .
\end{align}
To have a canonical kinetic term for the {\sf SU(5)} Higgs fields,
we normalize the fields by
\begin{align}
  \phi_{\bar c_j} & = \sqrt{2}\ {\sf 5}_{Hj} \ , & 
  \phi_{c_j} = \sqrt{2}\ {\sf 5}^*_{Hj} \ .
\end{align}
Now we are able to express the basic Yukawa couplings,
\begin{align} \label{eq:yukawa10}
  {\sf 16\ 16\ 10}_H %
  = \braket{\Psi_+^*}{B\, \Gamma_\mu}{\Psi_+} \phi_\mu \ ,
\end{align}
in {\sf SU(5)} fields.  The matrix $B$ is the equivalent of the charge
conjugation matrix ${\cal C}$ (which is dropped here) for {\sf
  SO(10)}.  We find
\begin{align} \label{eq:so10-w10}
  \begin{split}
    W_Y^{10} & = \sqrt{2} i f_{ab} \left[ - \left( {\sf 1}_a\; {\sf
          5}^*_b + {\sf 5}^*_a\; {\sf 1}_b \right) {\sf 5}_H \right.
    \\
    & \quad \left. + \left( {\sf 10}_a\; {\sf 5}^*_b + {\sf 5}^*_a\;
        {\sf 10}_b \right) {\sf 5}^*_H + \tfrac{1}{4} {\sf 10}_a\;
      {\sf 10}_b\; {\sf 5}_H \right] .
  \end{split}
\end{align}
Additionally to the known {\sf SU(5)} couplings (which are symmetric
now), we observe the couplings for the neutrinos, leading to their
Dirac masses $m_\nu^\text{\sc d}$, with $m_\nu^\text{\sc d}=m_u$.

\medskip

%====================================================================

Next, let us turn to ${\sf 16\ 16\ 10}_H\, {\sf 45}_H$. It appears in
four different invariants,
\begin{subequations} \label{eq:invariants}
  \begin{align} \label{eq:invariants-a}
    \left( {\sf 16\; 16} \right)_{10} 
    & \left( {\sf 10}_H\, {\sf 45}_H \right)_{10} , 
    \\ 
    \label{eq:invariants-b}
    \left( {\sf 16\; 16} \right)_{120} & \left( {\sf 10}_H\, {\sf
        45}_H \right)_{120} ,
    \\
    \label{eq:invariants-c} \left( {\sf 16\; 10}_H
    \right)_{16^*} & \left( {\sf 16\; 45}_H \right)_{16} ,
    \\
    \label{eq:invariants-d} \left( {\sf 16\; 10}_H \right)_{144^*} &
    \left( {\sf 16\; 45}_H \right)_{144} .
  \end{align}
\end{subequations}
Note that in Ref.~\cite{chang04}, only the second term
(\ref{eq:invariants-b}) is studied.
To calculate the different couplings, we generalize
Eqn.~(\ref{eq:so10-10H}) so that \cite{nath-syed}
\begin{align} \label{eq:so10-sigma45}
  {\sf 45} & : \!\!\! & \Sigma_{\mu\nu}\, \phi_{\mu\nu} & = -i \left(
    \bt{i}\bt{j}\,\phi_{c_i c_j} + b_i b_j\,\phi_{\bar{c}_i \bar{c}_j}
  \right.
  \nonumber \\
  & & & \qquad + \left. 2\, \bt{i} b_j\, \phi_{c_i \bar{c}_j} -
    \phi_{c_n \bar{c}_n} \right) ,
  \\
  {\sf 120} & : \!\!\!\!\! & \Gamma_\mu \Gamma_\nu \Gamma_\lambda
  \phi_{\mu\nu\lambda} & = \bt{i}\bt{j}\bt{k}\, \phi_{c_i c_j c_k} +
  b_i b_j b_k\, \phi_{\bar{c}_i \bar{c}_j \bar{c}_k}
  \nonumber \\
  & & & + 3\, \bt{i} b_j b_k\, \phi_{c_i \bar{c}_j \bar{c}_k} + 3\,
  \bt{i}\bt{j} b_k\, \phi_{c_i c_j \bar{c}_k}
  \nonumber \\
  & & & \quad + b_i\, \phi_{\bar{c}_n c_n \bar{c}_i} + \bt{i}\,
  \phi_{\bar{c}_n c_n c_i} \ .
\end{align}
The tensors of $\phi_{\mu\nu}$  can be decomposed into their
irreducible forms as 
\begin{align}
  \phi_{c_n\,\bar c_n} & = h \ , & \phi_{c_i\,c_j} & = h^{ij} \ ,
  \nonumber
  \\
  \phi_{\bar c_i\,\bar c_j} & = h_{ij} & \phi_{c_i\,\bar c_j} & =
  h^i_j + \frac{1}{5}\,\delta^i_j\,h \ .
\end{align}
with the ${\sf 1,\ 10^*,\ 10 \ \text{and}\ 24}$-plet,
normalized as
\begin{align}
  h & = \sqrt{10}\,H \ , & h_{ij} & = \sqrt{2}\,H_{ij} \ , \nonumber
  \\
  h^{ij} & = \sqrt{2}\,H^{ij} \ , & h^i_j & = \sqrt{2}\,H^i_j .
\end{align}
Analogously, we have for $\phi_{\mu\nu\lambda}$
\begin{align}
  \begin{split}
    \phi_{c_i\,c_j\,\bar c_k} & = f^{ij}_{k} + \tfrac{1}{4} \left(
      \delta^i_k\,f^j - \delta^j_k\,f^i \right) ,
    \\[1pt]
    \phi_{c_i\,\bar c_j\,\bar c_k} & = f^{i}_{jk} - \tfrac{1}{4}
    \left( \delta^i_j\,f_k - \delta^i_k\,f_j \right) ,
    \\[4pt]
    \phi_{c_i\,c_j\,c_k} & = \epsilon^{ijklm}\,f_{lm} \; ,
    \mspace{20mu} \phi_{\bar c_i\,\bar c_j\,\bar c_k} =
    \epsilon_{ijklm}\,f^{lm} ,
    \\[2pt]
    \phi_{\bar c_n\,c_n\,c_i} & = f^{i} , \mspace{80mu} \phi_{\bar
      c_n\,c_n\,\bar c_i} = f_{i} \ .
  \end{split}
\end{align}
We identify the ${\sf 5,\ 10,\ 45,\ 5^*,\ 10^*\ \text{and}\ 
  45^*}$-plet of {\sf 120}, which are normalized as
\begin{align}
  \begin{split}
    f^{i} & = \frac{4}{\sqrt{3}}\,h^{i} \; , 
    \quad f^{ij} = \frac{1}{\sqrt{3}}\,h^{ij} \; ,
    \quad f^{ij}_{k} = \frac{2}{\sqrt{3}}\,h^{ij}_{k} \ ,
    \\
    f_{i} & = \frac{4}{\sqrt{3}}\,h_{i} \ , 
    \quad f_{ij}  = \frac{1}{\sqrt{3}}\,h_{ij} \ , 
    \quad f^{i}_{jk} = \frac{2}{\sqrt{3}}\,h^{i}_{jk} \ .
  \end{split}
\end{align}

For the first invariant (\ref{eq:invariants-a}) we need the coupling
${\sf 10 - 10 - 45}$\,, which can be decomposed as
\begin{multline} \label{eq:so10-101045}
  \sqrt{2}\, \left[ \left( {\sf 5}_{10}\, {\sf 5}^*_{10} + {\sf
        5}_{10}^*\, {\sf 5}_{10} \right) {\sf 1}_{45} + {\sf 5}_{10}\,
    {\sf 5}_{10}\, {\sf 10}^*_{45} \right.
  \\
  + \left. {\sf 5}^*_{10}\, {\sf 5}^*_{10}\, {\sf 10}_{45} + \left(
      {\sf 5}_{10}\, {\sf 5}^*_{10} + {\sf 5}^*_{10}\, {\sf 5}_{10}
    \right) {\sf 24}_{45} \right] .
\end{multline}
Since the vev of the ${\sf 45}_H$ is taken in the {\sf 24}-direction
of {\sf SU(5)}, only the last two terms are relevant.  Now we
integrate out the heavy field {\sf 10} in Eqs. (\ref{eq:so10-w10},
\ref{eq:so10-101045}) by means of \mbox{$W_M^{10} = 2\ M_{10}\ {\sf
    5}\ {\sf 5}^*$} and obtain the coupling given in
Eqn.~(\ref{eq:so10-add-1}).

The calculation for the second term (\ref{eq:invariants-b}) is
straightforward.  We compute
\begin{align}
  W_Y^{120} & = \frac{i}{\sqrt{3}}\,f_{ab} \left[ \left( - {\sf 1}_a
      {\sf 10}_b + {\sf 10}_a {\sf 1}_b \right) {\sf 10}^*_H + 2\cdot
    {\sf 5}^*_a {\sf 5}^*_b {\sf 10}_H \right.
  \nonumber \\
  & \mspace{30mu} + 2 \left( {\sf 1}_a {\sf 5}^*_b - {\sf 5}^*_a {\sf
      1}_b \right) {\sf 5}_H + \left( {\sf 5}^*_a {\sf 10}_b - {\sf
      10}^*_a {\sf 5}^*_b \right) {\sf 5}^*_H
  \nonumber \\[4pt]
  &\mspace{30mu} - \left. \tfrac{1}{2}\, {\sf 10}_a {\sf 10}_b {\sf
      45}_H + \left( {\sf 5}^*_a {\sf 10}_b - {\sf 10}_a {\sf 5}^*_b
    \right) {\sf 45}^*_H \right]
\end{align}
and calculate the relevant terms of the coupling ${\sf 10 - 45 -
  120}$,
\begin{multline}
  \sqrt{3} \left[ 2 \left( {\sf 5}_{10}\;{\sf 24}_{45}\;{\sf
        45}_{120}^* + {\sf 5}_{10}^*\;{\sf 24}_{45}\;{\sf 45}_{120}
    \right) \right.
  \\
  - \left. {\sf 5}_{10}\;{\sf 24}_{45}\;{\sf 5}_{120}^* - {\sf
      5}_{10}^*\;{\sf 24}_{45}\;{\sf 5}_{120}
  \right] + \dots
\end{multline}
With the mass term 
\begin{align}  \label{eq:mass120}
  W_M^{120} & = M_{120} \left( \tfrac{1}{2}\; {\sf 10}\; {\sf 10}^* +
    {\sf 45}\;{\sf 45}^* - 2\cdot {\sf 5}\;{\sf 5}^* \right) ,
\end{align}
we then get the result of Eqn. (\ref{eq:so10-add-2}).

The remaining two operators read
\begin{align}
  \left( {\sf 16\; 10}_H \right)_{16^*} \left( {\sf 16\; 45}_H
  \right)_{16} & = \widetilde{\Psi} B\, \Gamma_\mu \phi_\mu
  \Sigma_{\nu\rho} \Psi\, \phi_{\nu\rho} \; ,
  \label{eq:so10-combinations16}
  \\
  \left( {\sf 16\, 10}_H \right)_{144^*} \left( {\sf 16\, 45}_H
  \right)_{144} & = \widetilde{\Psi} B\, \phi_\mu \Gamma_\nu \Psi\,
  \phi_{\mu\nu} \nonumber
  \\
  & \qquad - \left( \text{Eqn.~}(\ref{eq:so10-combinations16}) \right)
  .
  \label{eq:so10-combinations144}
\end{align}
The first expression in Eqn.~(\ref{eq:so10-combinations144}) describes
the reducible {\sf 160} representation. Since the {\sf 144} requires
\begin{align}
  \Gamma_\mu \widetilde\phi_\mu = 0 \ ,
\end{align}
we add $\Gamma_\mu^2 = \mathbbm{1}$ to project out the {\sf 16}
contribution which is already calculated in Eqn.
(\ref{eq:so10-combinations16}). Then we get the {\sf 144}
contribution just by the difference of the two terms.

Altogether, the couplings of the four operators read
\begin{subequations} \label{eq:so10-add}
  \begin{align} \label{eq:so10-add-1}
    \begin{split}
      \widehat Y_{10} & = \frac{h_{ij}^{10}}{M_{10}} \left\{
        \frac{1}{2}\, \epsilon_{abcde}\, \mathsf{10}_i^{ab}\,
        \mathsf{10}_j^{cd}\, \Sigma^e_f\, H^f \right.
      \\%[-1pt]
      & \mspace{70mu} - \left. \vphantom{\frac12} 2\, \overline{H}_a\,
        \Sigma^a_b \left( \mathsf{10}_i^{bc}\, \mathsf{5}^*_{jc} +
          \mathsf{10}_j^{bc}\, \mathsf{5}^*_{ic} \right) \right\}
    \end{split} \\[1pt]
    \begin{split}
      \widehat Y_{120} & = \frac{h_{ij}^{120}}{M_{120}} \left\{ - 2\,
        \epsilon_{abcde}\, \mathsf{10}_i^{ab}\, \mathsf{10}_j^{cf}\,
        H^d\, \Sigma^e_f \right.
      \\%[-1pt]
      & \mspace{78mu} - \overline{H}_a\, \Sigma^a_b \left(
        \mathsf{10}_i^{bc}\, \mathsf{5}^*_{jc} - \mathsf{10}_j^{bc}\,
        \mathsf{5}^*_{ic} \right)
      \\%[-1pt]
      & \mspace{72mu} \left. \vphantom{\frac12} - 4\, \overline{H}_a\,
        \Sigma_b^c \left( \mathsf{10}_i^{ab}\, \mathsf{5}^*_{jc} -
          \mathsf{10}_j^{ab}\, \mathsf{5}^*_{ic} \right) \right\}
      \label{eq:so10-add-2}
    \end{split} \\[1pt]
    \begin{split}
      \widehat Y_{16} & = \frac{h_{ij}^{16}}{M_{16}} \left\{
        \frac{1}{2}\, \epsilon_{abcde}\, \mathsf{10}_i^{ab}\,
        \mathsf{10}_j^{cf}\, H^d\, \Sigma^e_f \right.
      \\[0pt]
      & \mspace{50mu} + \left. \vphantom{\frac12} 2\, \overline{H}_a\,
        \Sigma^a_b\, \mathsf{10}_i^{bc}\, \mathsf{5}^*_{jc} -
        \overline{H}_a\, \mathsf{10}_i^{ab}\, \Sigma_b^c\,
        \mathsf{5}^*_{jc} \vphantom{frac12} \right\}
      \label{eq:so10-add-3}
    \end{split} \\[1pt]
    \begin{split}
      \widehat Y_{144} & = \frac{h_{ij}^{144}}{M_{144}} \left\{
        \epsilon_{abcde}\, \mathsf{10}_i^{ab}\, \mathsf{10}_j^{cd}\,
        \Sigma^e_f\, H^f \phantom{frac12} \vphantom{\frac12} \right.
      \\%[-1pt]
      & \mspace{70mu}- \tfrac{1}{2}\, \epsilon_{abcde}\,
      \mathsf{10}_i^{ab}\, \mathsf{10}_j^{cf}\, H^d\, \Sigma^e_f
      \\%[-1pt]
      & \mspace{35mu} \left. \vphantom{\frac12} + 2\, \overline{H}_a\,
        \Sigma^a_b\, \mathsf{10}_i^{bc}\, \mathsf{5}^*_{jc} +
        \overline{H}_a\, \mathsf{10}_i^{ab}\, \Sigma_b^c\,
        \mathsf{5}^*_{jc} \right\} ,
      \label{eq:so10-add-4}
    \end{split}
  \end{align}
\end{subequations}
where we only list the {\sf SU(5)} relevant terms.  
Without loss of generality, we can assume that the heavy particles all
have the same mass and can compare the couplings with those of {\sf
  SU(5)} (cf. Eqn.~(\ref{eq:consistent-w})),
\begin{subequations} \label{eq:so10-matrices-higher}
  \begin{align}
    f_1 & = \tfrac{1}{2} h^{10} + h^{144} \ ,
    \label{eq:so10-matrices-higher-f1}
    \\
    f_2 & = -2 \, h^{120} + \tfrac{1}{2} h^{16} - \tfrac{1}{2} h^{144}
    \ , \label{eq:so10-matrices-higher-f2}
    \\
    h_1 & = -2 \, h^{10} - h^{120} + 2\, h^{16} + 2\, h^{144}\ ,
    \label{eq:so10-matrices-higher-h1}
    \\
    h_2 & = -4 \, h^{120} - h^{16} + h^{144} \ .
    \label{eq:so10-matrices-higher-h2}
  \end{align}
\end{subequations}
Here, $h^{10}$ is symmetric whereas $h^{120}$ is antisymmetric;
$h^{16}$ and $h^{144}$, are not constrained by symmetry requirements.
We see that {\sf SO(10)} does not restrict the contributions from the
higher-dimensional operators, contrary to the basic Yukawa couplings
(cf. Eqn.~(\ref{eq:yukawa10})).  With these equations, we reduce the
{\sf SO(10)} case down to {\sf SU(5)}, so the implications of the
higher-dimensional operators for proton decay in {\sf SO(10)} are the
same as in {\sf SU(5)} \cite{cw03}.
%The other matrices, 

\medskip

If we consider the complete breaking of {\sf SO(10)} to ${\sf G_\sm}$,
more five-dimensional operators can appear.  {\sf SO(10)} can be
broken via a pair ${\sf 16}_H \oplus {\sf 16}^*_H$, where the {\sf
  SU(5)} singlet component obtains a vev ${\cal
  O}\left(M_\gut\right)$.  Then the two new dimension-five operators,
\mbox{${\sf 16\; 16\; 16}_H\, {\sf 16}_H$} and \mbox{${\sf 16\; 16\;
    16}^*_H\, {\sf 16}^*_H$}, generate Majorana masses for the
right-handed neutrinos.  If, moreover, the ${\sf 5}^*_{16}$ and ${\sf
  5}_{16^*}$ aquire vevs as well (as in Refs.~\cite{babu98}), these
operators allow additional contributions to
Eqn.~(\ref{eq:so10-matrices-higher}).  The second coupling was
partially worked out in Refs.~\cite{nath-syed}.
Alternatively, if we use a ${\sf 210}_H$ to break {\sf SO(10)},
additional terms can arise, since the {\sf 210} includes a ${\sf 24}$
of {\sf SU(5)} \cite{slansky81}.

\bigskip

%====================================================================

We extended the analysis of higher-dimensional operators in {\sf
  SU(5)} to {\sf SO(10)}.  In contrast to the basic Yukawa couplings,
these operators are not restricted compared to {\sf SU(5)}.  In the
simple case, where {\sf SO(10)} is broken to ${\sf G_\sm}$ via ${\sf
  16}_H$, ${\sf 16}^*_H$ and ${\sf 45}_H$ and the former have only
vevs in the {\sf SU(5)} singlet direction, these represent all
possible operators of dimension five.  Hence, it would be interesting
to study if this model, with only the Yukawa couplings ${\sf 16\ 16\ 
  10}_H$ and the dimension-five operators studied in this paper
(including those which generate Majorana masses for the right-handed
neutrinos), both reproduces the fermion masses and mixing angles and
is consistent with the experimental limit on proton decay.
\\[12pt]
{\slshape Acknowledgements}.  I am very grateful to W. Buchm\"uller
for the guidance in this project and to D. Emmanuel-Costa and A.
Utermann for fruitful discussions.  I thank S. Willenbrock for useful
comments on the manuscript.

\end{document}